\documentclass[a4paper,twocolumn,11pt]{quantumarticle}
\pdfoutput=1
\usepackage[utf8]{inputenc}
\usepackage[english]{babel}
\usepackage[T1]{fontenc}
\usepackage{amsmath}
\usepackage{hyperref}

\usepackage{tikz}
\usepackage{lipsum}

\usepackage{graphicx,mathtools,subcaption}
\usepackage{appendix,xcolor}
\usepackage[ruled,lined]{algorithm2e}

\graphicspath{{figs/}}

\begin{document}

\title{Unified framework for efficiently computable quantum circuits}

\author{Igor Ermakov}
\affiliation{Russian Quantum Center, Skolkovo, Moscow 143025, Russia.}
\affiliation{Department of Mathematical Methods for Quantum Technologies, Steklov Mathematical Institute of Russian Academy of Sciences, 8 Gubkina St., Moscow 119991, Russia.}

\author{Oleg Lychkovskiy}
\affiliation{Skolkovo Institute of Science and Technology, Bolshoy Boulevard 30, bld. 1, Moscow 121205, Russia.}
\affiliation{Russian Quantum Center, Skolkovo, Moscow 143025, Russia.}
\affiliation{Department of Mathematical Methods for Quantum Technologies, Steklov Mathematical Institute of Russian Academy of Sciences, 8 Gubkina St., Moscow 119991, Russia.}

\author{Tim Byrnes}
\email[]{corresponding author: tim.byrnes@nyu.edu}
\affiliation{New York University Shanghai, NYU-ECNU Institute of Physics at NYU Shanghai, Shanghai Frontiers Science Center of Artificial Intelligence and Deep Learning, 567 West Yangsi Road, Shanghai, 200126, China.}
\affiliation{State Key Laboratory of Precision Spectroscopy, School of Physical and Material Sciences, East China Normal University, Shanghai 200062, China}
\affiliation{Center for Quantum and Topological Systems (CQTS), NYUAD Research Institute, New York University Abu Dhabi, UAE.}
\affiliation{Department of Physics, New York University, New York, NY 10003, USA}

\maketitle

\begin{abstract}
Quantum circuits consisting of Clifford and matchgates are two classes of circuits that are known to be efficiently simulatable on a classical computer.  We introduce a unified framework that shows in a transparent way the special structure that allows these circuits can be efficiently simulatable.  The approach involves analyzing the transformation of operators in the Heisenberg picture, and viewing this as a spread within a network of basis operators. {\color{black} The operator amplitudes are found to follow a single variable Porter-Thomas distribution  for random universal quantum circuits.}  {\color{black} The number of operators with amplitude above a threshold value is shown to have a characteristic form} involving an initial exponential growth, saturation, then exponential decay in the presence of decoherence. {\color{black} We show the number of significant operators has a natural interpretation as the complexity of an numerical algorithm where errors can be consistently controlled as a function of the complexity of the simulation. }
\end{abstract}

\section{Introduction}

Calculating the time dynamics of a quantum circuit on a classical computer is in general a difficult task \cite{feynman2018simulating,aaronson2011computational,bremner2011classical,bremner2017achieving,arute2019quantum,boixo2018characterizing,bouland2019complexity,movassagh2023hardness}. The existence of quantum algorithms with a computational speedup over classical computers, already suggests that achieving similar performance classically is non-trivial \cite{nielsen2002quantum,shor1994algorithms,grover1996fast,harrow2009quantum,mosca2008quantum,tessler2017bitcoin}.  Viewing the register of the quantum computer as an interacting multiqubit system, this difficulty is closely related to the difficulty of calculating time dynamics of generic quantum many-body problems \cite{lloyd1996universal,eisert2015quantum,smith2019simulating,bluvstein2021controlling,schindler2013quantum}. Even for exactly solvable models whose equilibrium and near-equilibrium properties can be effectively computed \cite{gohmann2004integral,klumper1993thermodynamics,coleman1975quantum,gamayun2018impact} (usually, by means of the Bethe ansatz \cite{bethe1931theorie,korepin1997quantum,baxter2016exactly,das1989integrable}), exact far-from-equilibrium dynamics  for sufficiently large system sizes is typically out of reach.  In a quantum computing context, finding quantum computational advantage is typically based on this difficulty to simulate time dynamics classically  \cite{preskill2012quantum,lund2017quantum,hangleiter2023computational,arute2019quantum,zhong2020quantum,wu2021strong,madsen2022quantum,zhu2022quantum}.  In the seminal result by the Google collaboration, a quantum random circuit was used to show that equivalent results would be difficult to obtain using classical simulation \cite{arute2019quantum}.  In boson sampling, single photons enter a linear optical network, and the photonic detection probabilities are measured \cite{aaronson2011computational,broome2013photonic,spring2013boson,tillmann2013experimental}; a task that is computationally difficult to simulate.  In Gaussian boson sampling, single photons are replaced with squeezed light \cite{hamilton2017gaussian,zhong2020quantum}, where again the output photonic probabilities are difficult to predict classically.  In all these examples, the sequence of time evolutions of the quantum system lacking any simplifying symmetry makes classical prediction of the output a difficult task \cite{bouland2019complexity}. Quantum advantage is manifest by showing that the quantum computer outperforms the best available classical computers running the best algorithms. 

For qubit-based quantum circuits there are two notable exceptions to this.  The first is the foundational result of Gottesman and Knill \cite{gottesman1998heisenberg,aaronson2004improved}, which states that any quantum circuit that consist only of Clifford gates can be efficiently computed classically.  The second consist of circuits involving matchgates \cite{valiant2002expressiveness}, which are defined as any unitary operation of the form
\begin{align}
U_M = \left(
\begin{array}{cccc}
a_{00} & 0 & 0 & a_{01} \\
0 & b_{00} & b_{01} & 0 \\
0 & b_{10} & b_{11} & 0 \\
a_{10} & 0 & 0 & a_{11} \\
\end{array}
\right) ,
\label{matchgate}
\end{align}
where $a_{ij}, b_{ij} $ with $ i,j \in \{0,1 \} $ are matrix elements of a $2\times 2$ unitary matrix with the same determinant.  For a circuit consisting purely of matchgates on nearest neighbor qubits in a one dimensional geometry, and starting from a input state that is a product state, the expectation values in the computational basis can be calculated efficiently \cite{valiant2001quantum,valiant2002expressiveness,jozsa2008matchgates,brod2016efficient}.

A natural question that occurs here is whether there is any connection between such efficiently simulatable circuits. This is potentially an important question as better understanding the structure of such efficiently simulatable circuits may provide additional tools to discover other such circuits. Even for circuits which are not efficiently simulatable in the general case, in certain circumstances the complexity may only grow slowly, making them tractable in certain regimes. 
{\color{black} Another related question is to understand the complexity of a quantum circuit, particularly when decoherence is involved.  
Recently, it has become apparent that the complexity of simulating quantum dynamics is tied to operator growth. During the evolution, initially local operator transforms into a superposition of numerous non-local operators, thereby increasing the computational power needed to track them. Recent studies have explored universal patterns of operator growth, linking it with out-of-time-order correlators and information scrambling in various systems \cite{nahum2018operator,von2018operator,parker2019universal,mi2021information,landsman2019verified}. It has been found that in open systems, dissipation competes with scrambling by restricting operator growth \cite{khemani2018operator,zhang2019information,touil2021information,yoshida2019disentangling,Schuster_2023_Operator,Rakovszky_2022_Dissipation-assisted}. Advances in techniques in calculating the output of random quantum circuits in the presence of such imperfections have shown that this can reduce the computational complexity of simulating the output considerably \cite{aharonov2023polynomial,pashayan2015estimating,shi2022effect}. }

In this paper, we study this question and show a unified framework to understand both Clifford, matchgate, and other circuits which become effectively tractable. Working in the Heisenberg picture, we show that there is a universal type of behavior that is common to all quantum circuits in terms of the number of operators with significant amplitude. {\color{black} Our framework also allows a simple way to include decoherence, which also can be seen to directly affect the computability of the final result. We show that in random quantum circuits the coefficients follow a characteristic single-variable Porter-Thomas distribution with random signs. 
By introducing an explicit numerical algorithm to simulate observables of a quantum circuit, we show that the number of significant operators has the interpretation as the complexity of the quantum circuit.  }


\section{Heisenberg evolution}

\begin{figure}[t]
\includegraphics[width=\linewidth]{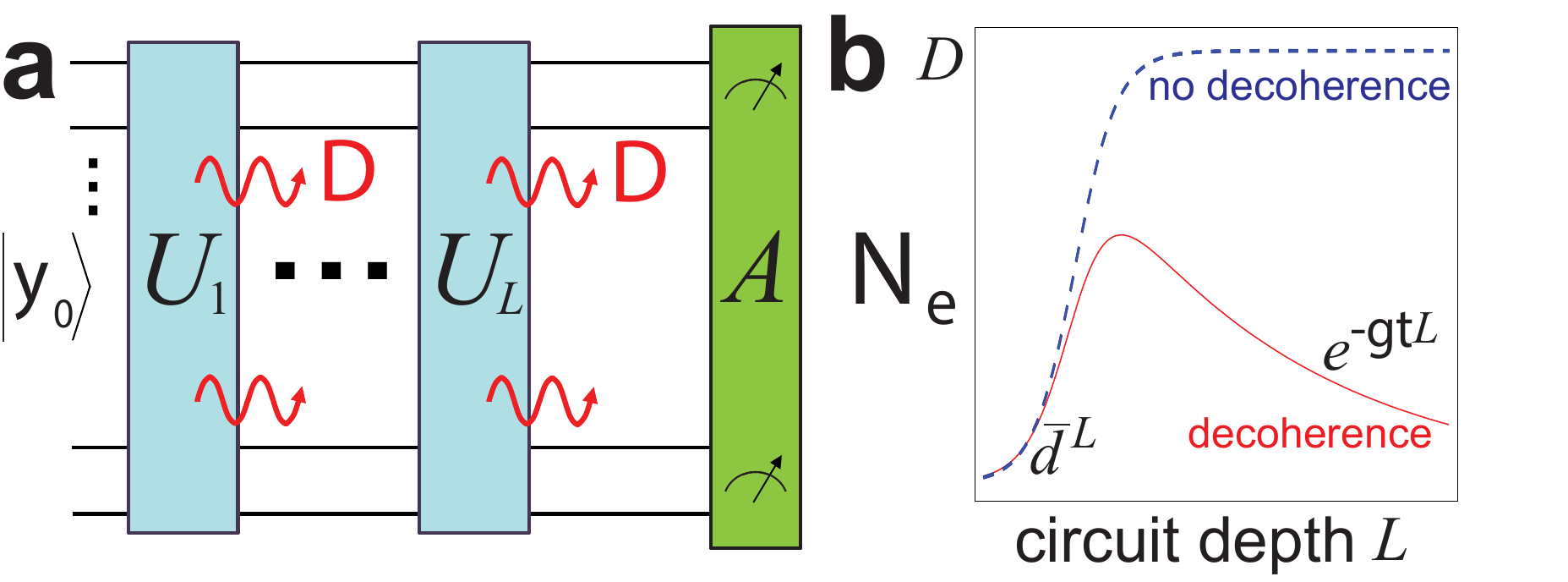}
\caption{(a) The problem considered in this paper.  A sequence of $ L $ unitary gates $ U_l $ acts on an initial state $ | \psi_0 \rangle $.  At the end, an observable $ A $ is measured.  During each gate, some decoherence $ \cal D $ occurs, indicated by the wavy lines.  (b) The generic evolution of the number of significant operators as defined in (\ref{nosigops}).  In the absence of decoherence (dashed line), the number of operators increases exponentially and saturates.  With decoherence, the number of operators initially increases exponentially but peaks and decays. \label{fig1}  }
\end{figure}

Consider an $ N $-qubit quantum circuit as shown in Fig. \ref{fig1}. The quantum circuit consists of a sequence of quantum gates consisting of $ L $ layers of unitaries $ U_l $ with $ l \in [1,L]$.  Each of the layers consist of a combination of gates that are choosable from a set $ \{ u_i \} $, where $ i $ labels all the possible gates that can be applied.  For example, in the case of a Clifford circuit, this would be the Hadamard, $ S = e^{iZ \pi/4} $ phase, and CNOT gates, where $ Z $ is the Pauli matrix.  

Suppose now we are interested in a particular observable  $ A $ at the end of the circuit, such that we wish to know what $ \langle A \rangle $ is.  Expand $ A $ in a complete orthonormal operator basis $ B_k $
\begin{align}
   {A} = \sum_{k=1}^{4^N} \lambda_k B_k . 
   \label{observableexpansion}
\end{align}
The $ \lambda_k $ are coefficients that may be computed by using the fact that $ \text{Tr} ( B_j^\dagger B_k ) = \delta_{jk} $, such that $ \lambda_k = \text{Tr} ( B_j^\dagger A ) $.  A simple choice of an orthonormal operator basis are Pauli strings, i.e. tensor products of Pauli operators $ B_k = \otimes_{n=1}^N P_n/\sqrt{2^N}  $ where $ P_n \in \{ I_n, X_n, Y_n, Z_n \} $. 

First, consider a quantum circuit consisting of a single layer $ U $, such that the gate depth is $L = 1$.  The observable evolves in the Heisenberg picture as 
\begin{align}
{A} & \rightarrow {A}' =  U^\dagger {A} U  = \sum_{k=1}^{4^N} \lambda_k B_k'
\label{basisevol}
\end{align}
Here we defined the evolved basis operator for the unitary $ U $ as
\begin{align}
 B_k' &  = U^\dagger B_k U  = \sum_{j=1}^{4^N} \Omega_{jk} B_j
 \label{newbasis}
\end{align}
and expanded it using the operator basis with coefficients 
\begin{align}
\Omega_{jk} = \text{Tr} (B_j^\dagger U^\dagger B_k U) . \label{omegadef}
\end{align}

{\color{black} For purely unitary operations and Hermitian operators $ B_j $, this is an orthogonal matrix satisfying
\begin{align}
\Omega^T \Omega = \Omega \Omega^T = I ,
\label{orthogonalmat}
\end{align}
as shown in the Appendix \ref{app:a}.  As such it is a real valued square matrix with columns and rows which are orthonormal vectors. }

Using this form, we see that the observable can be written as 
\begin{align}
{A}' = \sum_{j=1}^{4^N}  \lambda_j' B_j ,
\label{evolvedexp}
\end{align}
where the new coefficients are
\begin{align}
 \lambda_j' = \sum_{k=1}^{4^N} \Omega_{jk} \lambda_k .
\end{align}
Repeating this process for $ L $ gates as shown in Fig. \ref{fig1}, we may obtain the final coefficients
\begin{align}
\lambda_j^{(L)} = \sum_{k_1,\dots,k_L=1}^{4^N} 
\Omega_{j k_{1} }^{(1)} 
\Omega_{k_{1} k_{2} }^{(2)} 
\dots 
\Omega_{k_{L-1} k_L}^{(L)} \lambda_{k_L} ,
\label{manygatelambda}
\end{align}
{\color{black} and the operator after the $ L $ layers of the circuit is
\begin{align}
A^{(L)} = \sum_j \lambda^{(L)}_j B_j  . 
\label{aldef}
\end{align}
Equation (\ref{manygatelambda}) } takes the form of a sequence of $L $ {\color{black} orthogonal} matrix multiplications each of dimension $ 4^N \times 4^N $. {\color{black}  As such, for the purely unitary case (e.g. zero decoherence) the coefficients retain their normalization 
\begin{align}
\sum_j (\lambda_{j}^{(L)})^2 = \sum_k \lambda_{k}^2 . 
\end{align}
Throughout this paper we assume an observable $ A $ which has a normalization $\sum_k \lambda_{k}^2 = 1$ unless stated otherwise.  }

In Eqs. (\ref{basisevol}) and (\ref{newbasis}) above, we only considered unitary evolution and did not take into account the decoherence. Taking into account of decoherence does not change the overall procedure, and may be written in the same formalism, where the $ \Omega $ matrices also includes the effect of decoherence. {\color{black}  This is done by solving the equations of motion for the operators including decoherence \cite{teretenkov2023exact}, and again expanding the result in the operator basis. In this case the $ \Omega $ matrices are no longer orthogonal matrices, although they remain real valued. } 

We will generally consider a model of decoherence that is the same for all $ L $ layers (see Fig. \ref{fig1}(a)).  Implicit in this structure is that the total amount of applied decoherence is proportional to the number of layers, which is typically the scenario in many quantum computing contexts as deep circuits take a longer evolution time.  We give a simplified example of this below.  {\color{black} We consider a quantum computer without error correction implemented in the Noisy Intermediate Scale Quantum (NISQ) regime \cite{preskill2018quantum,chen2023complexity}. }

\section{Network representation}

\begin{figure}[t]
\includegraphics[width=\linewidth]{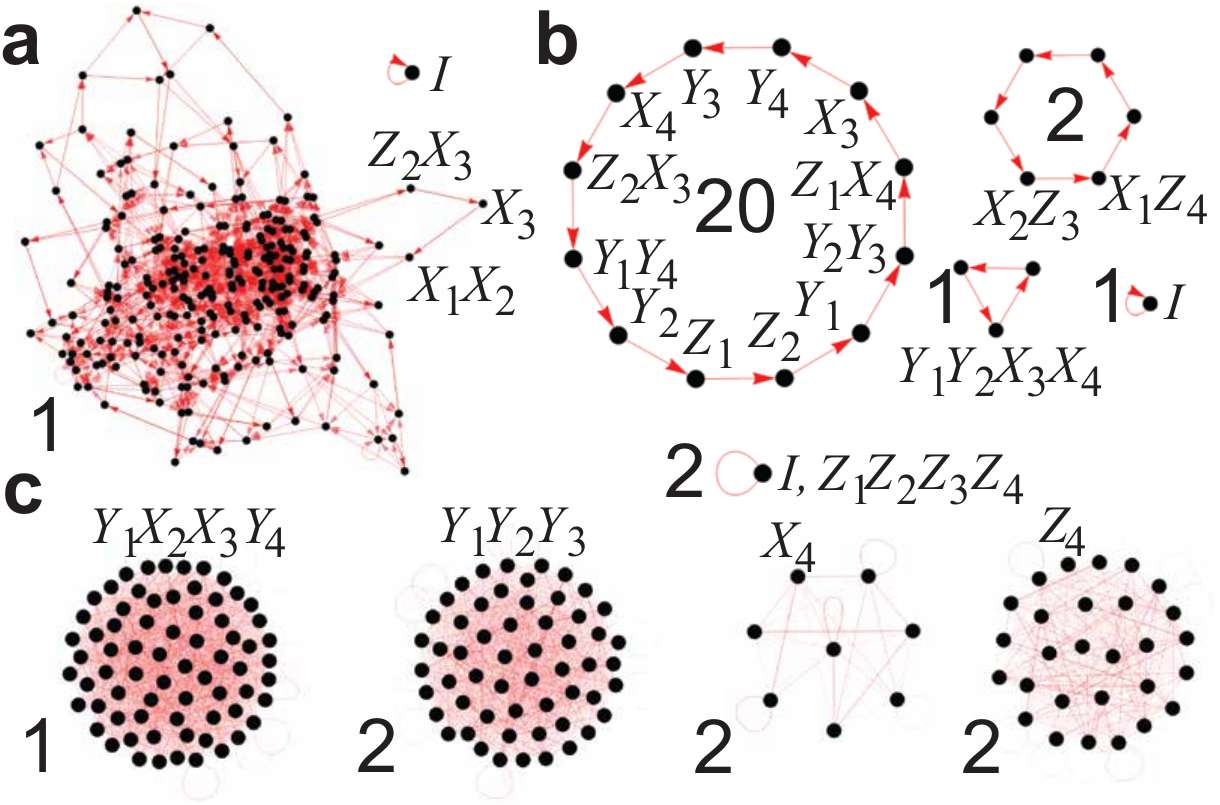}
\caption{Graphs corresponding to the adjacency matrix $ |\Omega_{jk} | $.  For each case we consider a $ N = 4 $ qubit circuit with circuit depth $ L = 4 $.  The gate sets chosen are (a) universal Clifford and $ T= e^{i Z \pi/8 } $ gate; (b) Clifford; and (c) matchgate circuit.  For (a)(b), each unitary $ U_l $ corresponds to a randomly selected single qubit and two qubit gate (including identity gates).  Each graph node corresponds to a Pauli string, some of which are labeled.  Large numbers indicate multiplicities of graphs with the same number of nodes and similar graphs (disregarding edge weights).  The opacity of the edges are in proportion to the absolute value of the weight $ | \Omega_{lk} | $.  
\label{fig2}}
\end{figure}

The product of the matrices $ \Omega = \Omega^{(1)} \Omega^{(2)} \dots \Omega^{(L)} $ contain all the information of how operators transform under a sequence of unitary evolutions.  It is illuminating to see the structure of these transformations for several examples of quantum circuits.  In Fig. \ref{fig2}, we visualize the matrix structure by interpreting $ | \Omega_{jk} |  $ as elements of an adjacency matrix that encodes a directed graph. Here, the nodes of the graphs represent the $ 4^N $ basis operators $ B_k $, which we take as Pauli strings.  The edges of the graphs represent the transformations that occur due to the action of the quantum circuit. In the case of the universal gate set of Clifford and $ T $ gates (Fig. \ref{fig2}(a)), we see a single connected graph (except the identity operator, which always appears as an isolated vertex), and no simple symmetry is apparent.  This is in fact an example of a relatively sparse graph, with a mean vertex out-degree $ \bar{d} \approx 7.3 $ for the case shown, and highest out-degree vertex $ d_{\max} = 16 $. For a completely random unitary, one would more typically have a fully connected graph, where $ \bar{d} \sim 4^N $.  For such a fully connected graph, the exponential number of operators generally means that the numerical simulation of the circuit is a difficult task, for this choice of basis operators.  Meanwhile, for the case shown in Fig. \ref{fig2}(a), the sparsity suggests that some observables may be simulated effectively. For example, the observable $ X_3 $  transforms to the single operator $ X_1 X_2 $ for the example shown.  As the depth of the circuit is increased, the mean vertex degree increases, and the graph approaches a maximally connected graph, increasing the complexity of the circuit.   

Contrast this to the graphs corresponding to a randomly chosen Clifford circuit as shown in Fig. \ref{fig2}(b). Clifford circuits simplify the graph structure considerably, such that they only consist of a single cycle, since Pauli strings transform to other Pauli strings.  The mean vertex out-degree in this case is $ \bar{d} = 1 $, which is the key aspect for simulatability of Clifford circuits. As long as one may keep track of the operator transformations at each step of the quantum circuit, one is able to compute the final full operator transformation.  Unlike the universal gate circuit of Fig. \ref{fig2}(a) where the graph increases in connectivity, the single cycle structure remains independent of the depth $ L $. 

Figure \ref{fig2}(c) shows the graphs corresponding to a matchgate circuit. In this case, the simplifying aspect is that there are several disconnected graphs, such that the graph breaks into components (in addition to the trivial identity operator).  This means that the matrix $ \Omega $ has a block diagonal structure. For each component, the vertices are fully connected, making them complete subgraphs.  The right-most graph contains $N^2$ vertices corresponding to operators known as Onsager strings, including  all operators $ Z_n $, $ n \in [1,N]$  \cite{Znidaric_2010_Exact,Shibata_2019_Dissipative_Kitaev_chain,Essler_2020_Integrability,teretenkov2023exact}.  This allows for an efficient computation of the Heisenberg evolution of the operators due to the block diagonal structure of the evolution which avoids the full $ 4^N$ operator space.

%


The network picture provides a basis for understanding the computational difficulty of a given quantum circuit in the Heisenberg picture.  Given an underlying graph structure of a quantum circuit, the observable transforms following graph edges and spreads throughout the network. 
Each successive application of a quantum gate then spreads through the network, such that more and more operators that must be kept account of, analogous to the spread of an epidemic \cite{qi2019quantum}.  If the number of operators that must be accounted for increases exponentially, the simulation becomes intractable.  

\section{Operator growth and decoherence}

The graphs shown in Fig. \ref{fig2} correspond to operator transformations of the full quantum circuit.  To see how the operators proliferate for deep quantum circuits, let us examine the spread of the operators throughout the network, layer by layer \cite{Parker_2019}.  It will also be interesting to include decoherence at this point. 
As a simplified model of decoherence, we apply decoherence after each layer of gates, such that unitary and decoherence operations occur alternately.  The general operator evolution under this type of evolution  is 
\small
\begin{align}
\lambda_j^{(L)} =
\sum_{k_1,\dots,k_L,k_1',\dots,k_L'=1}^{4^N} 
\Omega_{j k_{1} }^{(1)} 
{\cal D}_{k_1 k_1'}
\dots 
\Omega_{k_{L-1}' k_L}^{(L)} 
{\cal D}_{k_L k_L'}
\lambda_{k_L'} .
\label{manygatelambdawithdeco}
\end{align}
\normalsize
Here $ {\cal D}_{k k'} $ is the transformation associated with the decoherence process.  To consider a specific case, let us choose Lindblad-$ Z $ dephasing on each qubit with rate $ \gamma $.  For an operator basis that is a Pauli string, the basis operator evolves as $ B_k' = e^{-2\gamma q_k t} B_k $, where $ t $ is the decoherence time of a single layer, and $ q_k $ is the number of $ X $ or $ Y $ Pauli matrices involved in $ B_k $ \cite{teretenkov2023exact}.  The transformation matrix is therefore diagonal and has elements  $ {\cal D}_{k k'} = e^{-2 \gamma q_k t } \delta_{k k'} $.

An important quantity in the context {\color{black} of evolution} in the Heisenberg picture is the number of operators that have a significant amplitude.  We define an operator that has a significant amplitude as any coefficient in (\ref{evolvedexp}) that has an absolute value amplitude $ |\lambda_j^{(L)} | \ge \epsilon $, where $ \epsilon $ is a freely choosable cutoff parameter. The number of significant operators {\color{black} after the $ L $ layer circuit} then has a definition
\begin{align}
    {\cal N}_\epsilon = \sum_{j=1}^{4^N} T_\epsilon ( | \lambda_j^{(L)} |) ,
    \label{nosigops}
\end{align}
where $ T_\epsilon (x) = \theta(x-\epsilon) $ is the threshold function, which is a displaced Heaviside step function.  {\color{black} Later, we will propose that this quantity is related to the complexity of simulating the quantum circuit.  Due to the dependence on $ \epsilon $, clearly $ {\cal N}_\epsilon$ is not uniquely defined. We shall see that such a definition has a natural definition in terms of an explicit numerical algorithm, where operators with amplitude less than the cutoff can be systematically truncated at the expense of a loss in accuracy.  We will also see that while the numerical value of $ {\cal N}_\epsilon  $ will change with $ \epsilon $, it has a characteristic shape as the circuit depth and decoherence are changed, showing which regions of a quantum circuit are most difficult to simulate.  
}

\begin{figure}[t]
\includegraphics[width=\linewidth]{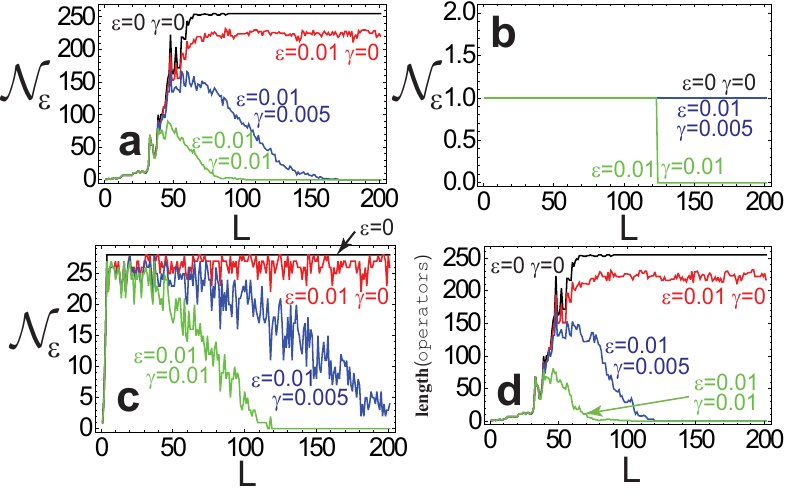}
\caption{(a)-(c) The number of significant operators for various quantum circuits with $ N = 4 $ as measured by the number of significant operators (\ref{nosigops}). Quantum circuits correspond to (a) Clifford + $ T $ gate; (b) Clifford; (c) matchgates.  The three lines marked by $ \gamma $ have a truncation parameter of $ \epsilon = 0.01 $.  The line marked as $ \epsilon = 0 $ is full number of operators with $ \gamma = 0 $ for comparison.  In each circuit, the gates are chosen randomly, but are kept consistent for different parameters $ \gamma, \epsilon$. {\color{black} For (a)(b), in each layer each qubit has a gate randomly chosen from $\{I, H, T \} $ or $\{I, H \} $ respectively, then between nearest neighbors a gate randomly chosen from $\{I, CZ \} $ is applied. For (c), in each layer a random matchgate (\ref{matchgate}) is applied between nearest neighbor qubits in sequence. (d) The number of operators, defined as the length of the \texttt{operators} list in Algorithm \ref{alg1} for the same quantum circuit as in (a) and the truncation and decoherence parameters as marked. } 
\label{fig3}  }
\end{figure}

Figure \ref{fig3}(a)-(c) shows $ {\cal N}_\epsilon $ for several types of random quantum circuits, each of depth $ L = 200 $ chosen as specified with various rates of decoherence $ \gamma $ and threshold $ \epsilon$.  Figure \ref{fig3}(a) shows a universal Clifford + $ T $ gate circuit. Initially, we see an exponential growth $\sim \bar{d}^L $ in the number of operators, consistent with the network epidemic picture.   In the untruncated case ($ \epsilon = 0 $) the growth in the number of operators occurs until it hits close to the maximum dimension $ 4^N $ of the operator space.  With truncation included, the number of operators is reduced, but remains constant without decoherence ($ \gamma = 0 $).  When decoherence is included, deep circuits reduce $ {\cal N}_\epsilon $ exponentially with $ L $.  {\color{black} Anticipating the interpretation of $ {\cal N}_\epsilon $ as a complexity,} we observe that the most difficult regions of simulation  should occur somewhere in the middle of a deep quantum circuit, where exponential increase in the number of operators has occurred but decoherence has not set in completely. 
{\color{black} Similar results have been shown recently using alternative methods \cite{yan2023limitations}. }

Figure \ref{fig3}(b) shows $ {\cal N}_\epsilon $ for a random Clifford circuit.  The number of operators always remains one, due to the average vertex out-degree $ \bar{d} = 1$, resulting in no exponential increase since $ \bar{d}^L = 1 $.  With the onset of decoherence, for deep enough circuits the single operator even becomes unnecessary to keep track of and $ {\cal N}_\epsilon $ drops to zero, indicating all expectation values are close to zero. Figure \ref{fig3}(c) shows the case for a matchgate circuit.  Here, the growth of $ {\cal N}_\epsilon $ is immediate due to the maximal connections in the graph of operators in the $ Z_L $ sector as seen in Fig. \ref{fig2}(c). This maximal value is however considerably smaller than in Fig. \ref{fig2}(a) due to the disconnected graph structure.  The number of significant operators then drops exponentially allowing for the possibility to truncate operators and thereby further save resources for deep circuits.  

{\color{black}
Using the network picture of operator growth, we can summarize the growth and decay of $ {\cal N}_\epsilon$ with the empirical relation}
\begin{align}
{\cal N}_\epsilon \propto \frac{D  e^{- \epsilon \gamma \tau L  } \bar{d}^L }{D-1+ a \bar{d}^{L}} ,
\label{nepsapprox}
\end{align}
where $ D $ is the order of the graph in question and $ \bar{d} $ is the mean vertex out-degree {\it per layer} characterizing the growth of the operators, and $ \tau, a $ are constants.  The typical form of this function is given in Fig. \ref{fig1}(b).  Initially there is an exponential increase in the complexity of the circuit.  Without decoherence, this increase eventually saturates, bounded by the order of graph, which is the full operator space dimension $ D = 4^N $ in the general case of a universal circuit, or at a lower level in the case symmetries are present as with matchgates.  When decoherence is present, there is an exponential decay, as various operators start to degrade at different rates.  There is thus a peak in the middle of the circuit, where the decoherence has not affected the circuit to a sufficient degree, but the exponential increase of operators has {\color{black} proliferated the operators}.  All the graphs of Fig. \ref{fig3}(a)-(c) agree with the relation (\ref{nepsapprox}).

{\color{black}
\section{Distribution of $ \lambda $ coefficients} 

In general, the distribution of the coefficients $ \lambda_j^{(L)} $ depends on the particular quantum circuit that is implemented.  For completely random quantum circuits without any structure, such as the universal Clifford + $ T $ gate circuit shown in Fig. \ref{fig3}(a), the probability a particular measurement outcome is known to follow the Porter-Thomas distribution \cite{porter1956fluctuations,boixo2018characterizing}.  This states that the probability of a measurement outcome 
$ |\langle n| U | \psi_0 \rangle|^2 $  having a value $ p $ follows the exponential distribution $ \text{Pr} (p) = D e^{-Dp} $.  Here $ | n \rangle $ is a state in the computational basis, and $ D $ is the dimension of the system.  

A similar argument can be made for the $ \lambda_j^{(L)}$ coefficients, with some modifications.  The primary difference of the case in question here is that the  coefficients $ \lambda_j^{(L)}$ are real, rather the complex for a quantum wavefunction.  This means that we must use the single-degree of freedom Porter-Thomas distribution \cite{porter1956fluctuations}. We derive this distribution explicitly (see Appendix \ref{app:b}) using random matrix theory methods and obtain for our case
\begin{align}
\text{Pr}(p) = \sqrt{\frac{D}{2 \pi \Lambda }} \frac{e^{-p D/(2\Lambda)}}{\sqrt{p } } , 
\label{ourptdist}
\end{align}
where $ \Lambda = \sum_{j} (\lambda_j^{(L)})^2 $ is the normalization of the coefficients.  Functionally, this has an additional factor of $ 1/\sqrt{p} $ which has the effect of making smaller amplitudes more likely.  

We verify the theoretical predictions with direct numerical analysis, where the coefficients $ (\lambda_j^{(L)})^2 $ are ordered from smallest to largest.  Using the distribution (\ref{ourptdist}) we may derive the form of the distribution of the ordered coefficients (see Appendix \ref{app:c})
\begin{align}
\lambda_{P(n)}^2 = \frac{2\Lambda}{D} (\text{erf}^{-1} \frac{n}{D} )^2 ,
\label{lambda2dist}
\end{align}
where $P(n) $ is a permutation function such as to reorder the coefficients from smallest to largest, labeled with $ n \in [1,D] $ and $ \text{erf}^{-1} (x) $  is the inverse of the Gauss error function. 
Figure \ref{fig4}(a) shows the numerically calculated $ \lambda_{P(n)}^2 $ for the same universal random Clifford and $ T $ gate circuit as in Fig. \ref{fig3}(a), for the zero decoherence case.  We see that the distribution of coefficients closely follows the distribution (\ref{lambda2dist}), confirming the predicted probability distribution (\ref{ourptdist}). We also compare it to the distribution based on the two-variable Porter-Thomas distribution $ \lambda^2_{P(n)} = -\frac{\Lambda}{D} \ln (1- n/D)$.  The numerically generated points agree better with the single variable counterpart (\ref{lambda2dist}), verifying that this is indeed the correct distribution.  

Figure \ref{fig4}(b) shows the distribution of $ \lambda_j^{(L)}$ including the sign, as well as showing the effect of decoherence, taken to be the same type as shown in Fig. \ref{fig3}.  We see approximately half the terms have a negative sign, as expected from a random quantum circuit.  Including decoherence generally reduces the magnitude of the coefficients without affecting the overall shape. This is precisely the effect as predicted by (\ref{lambda2dist}) where the distribution is the same up to the normalization of the coefficients.  Put another way, if the coefficients are renormalized as  $ \lambda^2_{P(n)}/\Lambda $, they follow a universal form independent of decoherence.  


\begin{figure}[t]
\includegraphics[width=\linewidth]{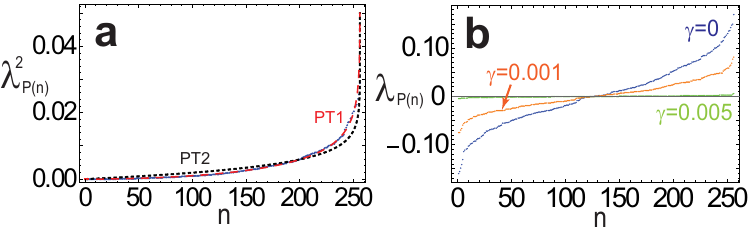}
\caption{\color{black} Distribution of coefficients $ \lambda_j^{(L)} $ for a random universal Clifford + $ T $ gate quantum circuits.  In (a) we evaluate the coefficients $ (\lambda_j^{(L)})^2 $ and reorder them from smallest to largest, with decoherence taken to be $ \gamma = 0 $ such that $ \Lambda = 1$ and $ L = 200$.   The dots are the values of $ (\lambda_j^{(L)})^2 $, and compared to   (\ref{lambda2dist}), labeled as PT1.  We also compare to the two variable Porter-Thomas distribution (\ref{lambdapt}), labeled as PT2.  In (b) the coefficients $ \lambda_j^{(L)} $ are ordered from smallest to largest taking into account of the sign. The coefficients taking into account of decoherence is also shown as marked.  
The parameters used are $ N = 4$, $ D = 4^4 = 256 $. \label{fig4}  }
\end{figure}

}

In general, amplitudes of long Pauli strings tend to be suppressed by the decoherence \cite{Rakovszky_2022_Dissipation-assisted,Wellnitz_2022_Rise,Schuster_2023_Operator,white2023effective}. To bound such suppression quantitatively,  we assume that only one- and two-qubit gates are used. Then  each unitary layer  changes the length of a Pauli string at most by one Pauli matrix. If we count the layers in the reverse order (from right to left), then there are at least  $(q_j+1-l)$   $X$ and $Y$ Pauli matrices in a Pauli string in the $l$th dissipative layer, adding the suppression factor $e^{-2\gamma (q_j+1-l) t}$.  Multiplying  these  factors with $l=1,2,\dots,q_j$ (we additionally assume that $q_j\leq L$), one gets the bound
\begin{equation}
|\lambda_j^{(L)}| \leq e^{-\gamma q_j(q_j-1) t}\, \|A\|.
\end{equation}

{\color{black}
\section{Numerical algorithm}

We now show that $ {\cal N}_\epsilon $ has a concrete interpretation as the complexity of simulating the quantum circuit by introducing an explicit numerical algorithm with complexity $ O({\cal N}_\epsilon) $.  The algorithm is shown in Algorithm \ref{alg1}.  The basic idea of the algorithm is to keep only operators with significant amplitude and truncate the remaining operators as they have a small amplitude. For simplicity, in Algorithm \ref{alg1} we assume that the observable to be measured is one of the Pauli strings $ B_m $ and the initial state is $| \psi_0 \rangle =  |0 \rangle^{\otimes N} $.   A key aspect of the algorithm is that it avoids explicitly storing computationally expensive $ \Omega$ matrices which have a dimension of $ 4^N \times 4^N $. Instead, an indexing system is used where the retained operators are referred to by a single number between $ m \in [1,4^N] $.  The lists \texttt{operators} and \texttt{lambda} then contain a list of the operators that are kept and their weights $ \lambda_j $.  

The algorithm proceeds step by step through the layers of the algorithm in reverse order (since we work in the Heisenberg picture).  For each operator that is kept, the effect of the unitary $ U_l $ and the decoherence is calculated which corresponds to the operator transformation $ B_k \rightarrow \sum_j \Omega_{jk} B_j $.  For circuits where the network connectivity is sparse (such as in Fig. \ref{fig2}(a)), the transformation of operators only spawns a relatively small number of operators for each layer of the circuit.  For example, in a universal Clifford and $ T $ gate circuit, a Pauli string maps at most to two other Pauli strings, e.g. $ T^\dagger X T = (X-Y)/\sqrt{2}$. The sparsity of the $\Omega$ matrix means that its entire contents does not need to be stored, reducing the computational overhead. 

Even with a sparse $\Omega$ matrix, multiple applications leads to an exponential growth of the number of operators as long as the columns contain more than one non-zero element.  Truncation is then performed by examining the coefficient of the new basis element $ B_j $, and is kept only if it has an amplitude larger than $ \epsilon $.  The parameter $ \epsilon $ plays a natural role as a variable that adjusts the accuracy of the simulation.  A smaller $ \epsilon $ keeps more operators and hence should give a more accurate result, at the cost of keeping more operators.  
The expectation value of the observable at the end of the circuit is 
\begin{align}
\langle A^{(L)} \rangle = \sum_j \lambda_j^{(L)} \langle \psi_0 | B_j | \psi_0 \rangle, 
\end{align}
where we used (\ref{aldef}).  In the case of the initial state being $ | \psi_0 \rangle = | 0 \rangle^{\otimes N} $, $ B_j $ operators that are diagonal (i.e. products of $ I $ and $ Z $ matrices) give a matrix element of 1, while non-diagonal $ B_j $ matrices give 0.  Hence simply summing the $ \lambda_j^{(L)} $ for the diagonal operators give the expectation value.

The complexity of the algorithm is directly proportional to the number of operators (i.e. the length of the \texttt{operators} list) that must be kept track of during the simulation. The number of operators that are kept is similar, but not precisely the same as $ {\cal N}_{\epsilon} $   (see Fig. \ref{fig3}(d)).  The reason is that $ {\cal N}_{\epsilon} $ is defined in a way such that the full set of all operators is calculated for a given layer, then the number of operators with an amplitude larger than $ \epsilon $ is counted (``global truncation'').  In Algorithm \ref{alg1}, the operator is dropped as soon as its amplitude drops below $ \epsilon $. In spread throughout the network of operators, this corresponds to ``pruning'' new operators in with small weights, as opposed to globally removing operators with small weights.  A difference between the two approaches arises because there is a possibility that the subsequent evolution of a dropped operator can contribute to operators which are not dropped, due to the presence of loops in the graphs as shown in Fig. \ref{fig2}.  For this reason the dominant effect of the pruning is to modify the $ \lambda_j $ weights, rather than change the number of operators kept. The main difference tends to occur when the number of operators is very small, where the significant operators form several disconnected graphs.  Below the percolation threshold, the number of operators diminish rapidly to zero.  The peak values however remain quite consistent and therefore the complexity of Algorithm \ref{alg1} is well-approximated by $ O( {\cal N}_{\epsilon}) $.

Figure \ref{fig5}(a)(c) shows the expectation value of $ Z_N $ for the same universal Clifford + $ T $ gate circuit as in Fig. \ref{fig3}(a), comparing the full calculation ($ \epsilon = 0 $) with Algorithm \ref{alg1} ($\epsilon = 0.01$) with truncation.   We see that there is excellent agreement for all times.  The error of the approximation is shown in Figs. \ref{fig3}(b)(d) for Algorithm \ref{alg1}, in comparison to the global truncation scheme.  We see that as expected, the global truncation performs better, since it more consistently truncates away operators with small amplitudes.  The global truncation scheme has a complexity of $ O(4^L) $ since it must initially evaluate the coefficients for the full set of operators, then truncate ones with small magnitude. In this sense, Algorithm \ref{alg1} is more efficient, at the cost of a reduced accuracy.  The magnitude of the error of the global truncation scheme is of the order of $ \epsilon $. This can be true despite the number of operators being discarded being rather large.  For example, in the case including decoherence, the number of operators kept rapidly reduces to less than half the full number (Fig. \ref{fig3}(a)), while the error remains at the level of $ \epsilon $.  The reason for the relatively insensitivity to the number of discarded operators is that the signs of the coefficients $ \lambda_j^{(L)} $ are typically randomly distributed, as seen in Fig. \ref{fig4}(b). In the case including decoherence, the number of operators retained rapidly diminishes as can be see in Fig. \ref{fig3}(a).  Hence for very deep circuits with decoherence it is not difficult to obtain high accuracy results, primarily because most operators have very small amplitudes.  

\begin{figure}[t]
\includegraphics[width=\linewidth]{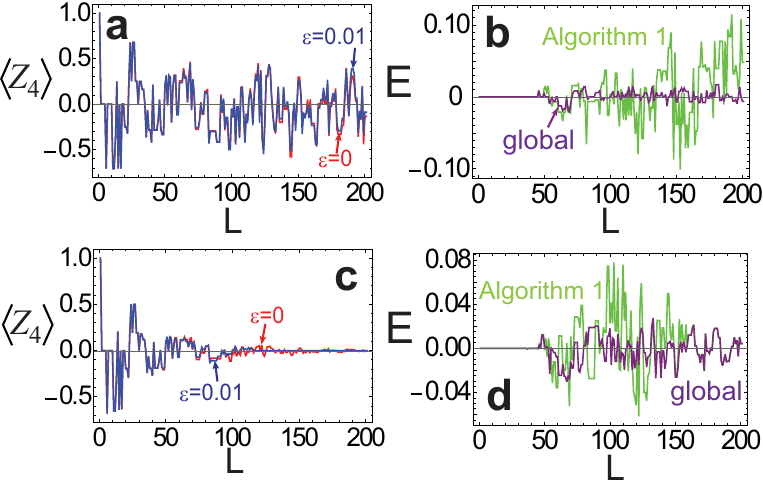}
\caption{\color{black} Evaluating expectation values using operator truncation.  (a)(c) Expectation value of the observable $ A = Z_4$ for the universal Clifford + $ T $ gate using Algorithm \ref{alg1} with no truncation ($\epsilon = 0 $) and with truncation ($ \epsilon = 0.01 $).  The decoherence values are (a) $ \gamma = 0 $ and (c) $ \gamma = 0.005 $. (b)(d) The error of the expectation defined as $ E = \langle A \rangle_{\epsilon}  - \langle A \rangle_{\epsilon=0}$.  Errors using Algorithm \ref{alg1} and the global truncation procedure is shown for $ \epsilon = 0.01 $ and decoherence (b) $ \gamma = 0 $ and (d) $ \gamma = 0.005 $.  Decoherence time is set to $ t = 1 $.  
\label{fig5}  }
\end{figure}

\begin{algorithm}
\caption{Approximate evaluation of an observable $ \langle A \rangle $.  \label{alg1} }
         \SetKwInOut{Input}{input}
        \SetKwInOut{Output}{output}
        \SetKwInOut{Assumptions}{assumptions}
        \SetKwComment{Comment}{/* }{ */}
        \SetArgSty{textup}
        \Input{Truncation parameter $ \epsilon $, transformation $ \Omega_{jk}^{(l)} $, observable $ A =B_m $}
        \Output{Expectation value $ \langle A \rangle $}
        \Assumptions{Initial state $ |\psi_0 \rangle = |0 \rangle^{\otimes N } $, $ A = B_m $.   }
    \SetKwBlock{Beginn}{beginn}{ende}
    \Begin{
\texttt{operators} $ \gets  [m] $ \Comment*[r]{List of length 1}
\texttt{lambda} $ \gets  [1.0]$\;  
        \For{$l \gets L$ to $1$}{
 \texttt{operatorsnew} $ \gets  []$  \Comment*[r]{Empty list}  
 \texttt{lambdanew} $ \gets  []$ \;
         \For{$n \gets 1$ to {\bf length}(\texttt{operators}) }{ 
                  $ \lambda_k  \gets \texttt{lambda}[n] $ \; 
         $ k  \gets \texttt{operators}[n] $ \;
     \For{$j$ such that $|\Omega_{jk}^{(l)} |>0$}{ 
  \If{$|\lambda_k \Omega_{jk}^{(l)}  | > \epsilon $}{
 {\bf append}(\texttt{operatorsnew}, $j$ ) \;
   {\bf append}(\texttt{lambdanew}, $\lambda_k \Omega_{jk}^{(l)}  $) \;
                }
                } 
        }
\texttt{operators} $ \gets $ \texttt{operatorsnew}\; 
\texttt{lambda} $ \gets $ \texttt{lambdanew}\; 
        }

  $ x \gets 0 $ \Comment*[r]{Compute expectation value}
  \For{$n \leq 1$ to {\bf length}(\texttt{operators}) }{ 
      \If{\texttt{operators}$[n]$ is diagonal operator}{
        $ x \gets x + \texttt{lambda}[n]$ \;
      }
  }  
        \Return $x$
    }
\end{algorithm}

}

%

\section{Basis choice}

The structure of the matrix $ \Omega $ is highly dependent upon the choice of basis operators $ B_k $.  Pauli strings are merely a convenient choice and other bases may be chosen, which potentially simplifies the graph structure. {\color{black} We now explicitly construct the optimal basis which reduces the graph structure to disconnected single vertices (an empty graph).  Consider a quantum circuit corresponding to the unitary $ U $.  We may write this as $ U = e^{-i H_\text{eff} t} $, where $ H_\text{eff}$ is the effective Hamiltonian that generates the unitary corresponding to the circuit.  Then the optimal choice of basis operators are
\begin{align}
B_j = \tilde{B}_{nm} = | E_n \rangle \langle E_m | 
\end{align}
where $ j= (n,m) $ is an index that combines the pair of energy labels and $| E_n \rangle $ are the eigenstates of $ H_\text{eff}$.  Then using the definition (\ref{omegadef}), we have
\begin{align}
\Omega_{jk} & = \text{Tr} ( \tilde{B}_{nm}^\dagger U^\dagger B_{n'm'} U ) \nonumber \\
& = \langle E_n | E_{n'} \rangle  \langle E_m | E_{m'} \rangle e^{-i(E_{m'} - E_{n'}) t } \nonumber \\
& = \delta_{jk} e^{-i(E_m - E_n) t }  ,  
\end{align}
which is a diagonal matrix.  This is equivalent to diagonalizing $ \Omega $ for a given quantum circuit and and working in the operator basis that diagonalizes the matrix. 

The implication of this is that by choosing an improved operator basis, it is possible to simplify the network, which in this case makes the network structure completely trivial. However, finding such an optimal basis is in general highly non-trivial, as it requires finding the eigenvectors of the quantum circuit.  Hence in the sense of the practical solution of the output of a quantum circuit, it may not be desirable to solve it in this way.  What may however be beneficial is by using an alternative basis, it partially simplifying the matrix $ \Omega $ may improve the performance considerably.  For example, performing local unitary transformations only may simplify the network such as to reduce the mean vertex degree, then an approximate numerical algoirthm such as Algorithm \ref{alg1} may be used to estimate expectation values.  

}


\section{Summary and conclusions}

We have shown that it is possible to understand several efficiently simulatable quantum circuits, specifically the Clifford and matchgate quantum circuits in terms of a single framework.  While conventionally these circuits are understood in terms of the stablizer formalism and fermionic transformations, our approach uses a simple Heisenberg picture of operators and analyzes the operator connectivity in a graph theoretical framework.  The quantum circuit can be equivalently thought of as a proliferation of operators through a network of basis operators. Eq. (\ref{nepsapprox}) captures the relevant ingredients that determine whether a particular observable is tractable or not. There is an interplay of the vertex out-degree, graph component structure, and decoherence which affects the simulatability of a circuit.   In the network picture, decoherence reduces the number of active nodes in the network, which helps to contain the spread to all operators. {\color{black} The distribution of the coefficients for a universal random circuit was derived to have the single variable Porter-Thomas distribution with randomly distributed signs.  In order to give a  interpretation of $ {\cal N}_\epsilon $ as a complexity, we introduced Algorithm \ref{alg1} which has a complexity $ O({\cal N}_\epsilon) $ where operators are truncated depending upon whether its expansion has a magnitude larger than $ \epsilon $.  The algorithm is relatively insensitive to the 
truncation due to the random distribution of signs, such that the coefficients below the cutoff tend to sum to zero.  The coefficients that are near to the cutoff, and any imbalance in their signs, contribute to the overall error. 

We introduced Algorithm \ref{alg1} as a way of giving $ {\cal N}_\epsilon $ a concrete interpretation as a complexity.  As such, the the algorithm can be further improved such that it can be robust computing engine for simulating quantum circuits. For example, the operator basis can be optimally chosen such as to reduce the operator proliferation as much as possible. A possible figure of merit here would be the mean vertex out-degree per layer $ \bar{d} $ such as to make the operator growth as slow as possible.  Another improvement is to temporarily keep more operators before immediately truncating them away.  Even with the rudimentary implementation that we demonstrated in Fig. \ref{fig5}, the algorithm shows good agreement despite only keeping a fraction of the total operator number, particularly when decoherence is involved.  A better understanding of the nature of efficiently computable circuits may lead to the discovery of other classes of simulatable circuits, and a better understanding of the realms of quantum computational advantage. 
}

\section*{Acknowledgments}
The authors thank Junheng Shi and Vsevolod Yashin for illuminating discussions. This work is supported by the National Natural Science Foundation of China (62071301); NYU-ECNU Institute of Physics at NYU Shanghai; Shanghai Frontiers Science Center of Artificial Intelligence and Deep Learning; the Joint Physics Research Institute Challenge Grant; the Science and Technology Commission of Shanghai Municipality (19XD1423000,22ZR1444600); the NYU Shanghai Boost Fund; the China Foreign Experts Program (G2021013002L); the NYU Shanghai Major-Grants Seed Fund; Tamkeen under the NYU Abu Dhabi Research Institute grant CG008; and the SMEC Scientific Research Innovation Project (2023ZKZD55). 

I. E. and O. L. are supported by Rosatom in the framework of the Roadmap for Quantum computing (Contract No. 868-1.3-15/15-2021 dated October 5,2021).




\providecommand{\noopsort}[1]{}\providecommand{\singleletter}[1]{#1}%

\onecolumn
\appendix

\section{Orthogonality of the $ \Omega $ matrix}
\label{app:a}

We show here that for purely unitary evolutions and Hermitian basis operators,  $ \Omega $ is an orthogonal matrix. 

First let us define the inner product in the space of operators as
\begin{align}
(A|B) = \text{Tr} (A^\dagger B) . 
\end{align}
For our basis operators we thus have $ (B_j | B_k ) = \delta_{jk} $.  The matrix elements of the $ \Omega $ matrix are then in this notation
\begin{align}
\Omega_{jk} = (B_j | U^\dagger B_k U ) . 
\end{align}

The matrix elements of the left hand side of (\ref{orthogonalmat}) are
\begin{align}
[ \Omega^T \Omega ]_{jl} & = \sum_k \Omega_{kj} \Omega_{kl} \nonumber \\
& = \sum_{k} ( B_k | U^\dagger B_j U ) (B_k | U^\dagger B_l U) \nonumber \\
& = \sum_{k} ( U^\dagger B_j^\dagger U | B_k^\dagger ) (B_k | U^\dagger B_l U) 
\label{orthoproof1}
\end{align}
where we used the fact that $ (A|B) = (B^\dagger | A^\dagger ) $ from the cyclic property of the trace.  Assuming that $ B_j^\dagger = B_j$, we then have
\begin{align}
[ \Omega^T \Omega ]_{jl} & = \sum_{k} ( U^\dagger B_j U | B_k ) (B_k | U^\dagger B_l U)  \nonumber \\
& = ( U^\dagger B_j U | U^\dagger B_l U)  \nonumber \\
& = \text{Tr} ( U^\dagger B_j^\dagger U  U^\dagger B_l U )   \nonumber \\
& = \text{Tr} (  B_j^\dagger  B_l  ) = \delta_{jl}
\label{orthoproof2}
\end{align}
which shows that $ \Omega^T \Omega = I $.

\section{Probability distribution of $ \lambda $ coefficients} 
\label{app:b}

We derive the probability distribution of the $ \lambda $ coefficients by adapting the two-variable Porter-Thomas distribution for random quantum circuits \cite{boixo2018characterizing} to the single variable case.  Our aim is to derive the probability that the coefficient $ \lambda_l$ following a random quantum circuit has a magnitude squared satisfying $ \lambda_l^2 = p \ge 0  $. We consider the $l $th coefficient without loss of generality, where $ l \in [1,D] $.  We take the coefficients to have a norm
\begin{align}
\sum_{k=1}^D \lambda_k^2 = \Lambda .  
\label{normdef}
\end{align}
For a normalized observable $ A $ and no decoherence we have $ \Lambda = 1 $.  Decoherence will typically have an effect to reduce $ \Lambda <1  $.  

The probability is given by 
\begin{align}
\text{Pr}(p) = \frac{\int_{-\infty}^{\infty} \prod_{j=1}^D d \lambda_j \delta( \sum_k \lambda_k^2 - \Lambda) \delta( \lambda_l^2 - p)}{\int_{-\infty}^{\infty} \prod_{j=1}^D d \lambda_j \delta( \sum_k \lambda_k^2 - \Lambda) } .  
\end{align}

The numerator is evaluated as
\begin{align}
& \int_{-\infty}^{\infty} \prod_j d \lambda_j \delta( \sum_k \lambda_k^2 - \Lambda) \delta( \lambda_l^2 - p) \nonumber 
= 
\int_{-\infty}^{\infty} \prod_j d \lambda_j \frac{1}{2 \pi} \int_{-\infty}^{\infty} dt e^{it(\sum_k \lambda_k^2 - \Lambda) } \frac{1}{2 \pi}  \int_{-\infty}^{\infty} dw e^{iw(\lambda_l^2 - p) }   \nonumber \\
& = \frac{1}{2 \pi} \int_{-\infty}^{\infty} dt e^{-it \Lambda} \left( \int_{-\infty}^{\infty} \prod_j d \lambda e^{it \lambda^2} \right)^{D-1} 
\frac{1}{2 \pi}  \int_{-\infty}^{\infty} dw e^{- iw p }  \int_{-\infty}^{\infty} \prod_j d \lambda_l e^{i(t+w) \lambda_l^2 }  \label{templine}  \\
& = \frac{1}{2 \pi} \int_{-\infty+ i \epsilon }^{\infty+ i \epsilon } dt e^{-it\Lambda} \sqrt{\frac{\pi}{-it}}^{D-1}   
  \frac{1}{2 \pi} \int_{-\infty+ i \epsilon }^{\infty+ i \epsilon } dw e^{- iw p } \sqrt{\frac{\pi}{-i(t+w) }} \label{templine2}\\
& = \frac{\sqrt{\pi}^{D-3} }{2  \sqrt{p} } \int_{-\infty+ i \epsilon }^{\infty+ i \epsilon } dt e^{-it(\Lambda-p) } \frac{1}{\sqrt{-it}^{D-1}} \label{templine3} = \frac{2^{\text{mod}_2 D-1 } \sqrt{\pi}^{D-1} \sqrt{\Lambda-p}^{D-3} }{2  \Gamma (\frac{D-1}{2} ) \sqrt{p} } . 
\end{align}
Here in (\ref{templine}) we used the fact that for $ \text{Im} (t) > 0 $, 
\begin{align}
\int_{-\infty}^{\infty} d \lambda e^{it \lambda^2 } =  \sqrt{\frac{\pi}{-it}} . 
\end{align}
In (\ref{templine2}) and (\ref{templine3})  we used the Fourier transforms
\begin{align}
\int d w e^{-iwp} \frac{1}{\sqrt{-i(t+w)}} = 2 \sqrt{\frac{\pi}{p}} e^{ipt} \nonumber \\
\int dt e^{-itx} \frac{1}{\sqrt{-it}^n}  = \frac{2^{\text{mod}_2 n } \pi \sqrt{x}^{n-2}}{\Gamma(n/2) } 
\end{align}
where $ \Gamma $ is the Gamma function and $ 2^{\text{mod}_2 n } $ is equal to 2 for odd $ n $ and 1 for even $ n $.  

The denominator is evaluated in a similar way
\begin{align}
& \int_{-\infty}^{\infty} \prod_{j=1}^D d \lambda_j \delta( \sum_k \lambda_k^2 - \Lambda) \nonumber \\
& = \frac{1}{2 \pi} \int_{-\infty}^{\infty} dt e^{-it\Lambda} \left( \int_{-\infty}^{\infty} \prod_j d \lambda e^{it \lambda^2} \right)^{D}  \nonumber \\
& = \frac{1}{2 \pi} \int_{-\infty}^{\infty} dt e^{-it \Lambda} \sqrt{\frac{\pi}{-it}}^{D}   \nonumber \\
& = \frac{2^{\text{mod}_2 D } \sqrt{\pi}^D \sqrt{\Lambda}^{D-2} }{2 \Gamma(\frac{D}{2} ) } .
\end{align}

Combining these we obtain the probability distribution
\begin{align}
\text{Pr}(p) & \propto \frac{\sqrt{1-p/\Lambda}^{D-3}}{\sqrt{p\Lambda}}  , 
\end{align}
where we have dropped all constant factors as we will normalize the probability distribution. For $ D/\Lambda \gg 1 $, we may approximate $ 1- p/\Lambda \approx e^{-p/\Lambda} $ and after normalization over the range $ p \in [0,1] $ we obtain 
\begin{align}
\text{Pr}(p) \approx \sqrt{\frac{D}{2 \pi \Lambda }} \frac{e^{-p D/(2\Lambda)}}{\sqrt{p } } . 
\label{porterthomaslike}
\end{align}

\section{Distribution of the ordered $ \lambda $ coefficients} 
\label{app:c}

Now consider a particular set of $ \lambda_j $ coefficients generated by a random quantum circuit.  Consider reordering the coefficients in order of $ \lambda_j^2 $, from smallest to largest. Let the reordering function be the permutation function $ P(n ) $, such that $ \lambda_{P( n)}^2 $ with $ n \in [1,D] $ gives the reordered function.  Here we derive the form of this distribution.  

Consider $ D$ draws from the probability distribution (\ref{porterthomaslike}).  This will be used to populate the coefficients $ \lambda_k $.  Of these, the number that has a magnitude between $ p $ and $ p + dp $, where $ p = \lambda^2 $ is the square of the coefficient,  will be
\begin{align}
d n = D \sqrt{\frac{D}{2\pi}} \frac{e^{- p D/(2\Lambda)}}{\sqrt{p \Lambda}} dp . 
\end{align}
The cumulative number of these up to the magnitude $ \lambda^2 $ is 
\begin{align}
n( \lambda^2) & = \int_0^{\lambda^2} dp D \sqrt{\frac{D}{2\pi}} \frac{e^{- p D/(2 \Lambda)}}{\sqrt{p \Lambda}} \nonumber \\
& = D \text{erf} \sqrt{\frac{ D \lambda^2}{2 \Lambda}} . 
\end{align}
Rearranging this in terms of $ \lambda^2 $ we have
\begin{align}
\lambda_{P(n)}^2 = \frac{2 \Lambda}{D} (  \text{erf}^{-1} \frac{n}{D})^2 .
\end{align}

In Fig. \ref{fig4} we wish to compare to the two-variable Porter-Thomas distribution. First, taking into account of the normalization (\ref{normdef}) we obtain the normalized distribution
\begin{align}
    \text{Pr} (p) = \frac{D}{\Lambda} e^{-D p/\Lambda} .
\end{align}
Assuming this distribution instead, the number with a magnitude between $ p $ and $ p + dp $ is
\begin{align}
dn = \frac{D^2}{\Lambda} e^{-D p /\Lambda} dp . 
\end{align}
The cumulative number up to a magnitude $ \lambda^2 $ is
\begin{align}
n(\lambda^2) & = \int_0^{\lambda^2} dp \frac{D^2}{\Lambda}  e^{-D p /\Lambda } \nonumber \\
& = D ( 1- e^{-D \lambda^2  /\Lambda }) . 
\end{align}
Rearranging, we have the distribution of coefficients assuming a Porter-Thomas distribution 
\begin{align}
\lambda^2_{P(n)} = - \frac{\Lambda}{D} \ln (1 - \frac{n}{D} ) .  
\label{lambdapt}
\end{align}
For random quantum circuits the ordered distribution of probability outcomes has been shown to agree with this distribution \cite{boixo2018characterizing,shi2022quantum}.

\end{document}